\documentclass{aa}
\usepackage{xcolor}
\usepackage{gensymb}
\usepackage[colorlinks,allcolors=blue]{hyperref}
\usepackage{comment}

\usepackage{graphicx}

\definecolor{cmt}{rgb}{0.5,0.0,0.0}
\definecolor{al}{rgb}{0.6,0.2,0.0}
\definecolor{mk}{rgb}{0.4,0.4,0.0}
\definecolor{no}{rgb}{0,0.0,0.4}
\definecolor{ns}{rgb}{0.4,0.0,0.0}
\definecolor{hr}{rgb}{0.4,0.0,0.0}

\newcommand{\todo}[1]{{\color{cmt}ToDo: #1}}

\newcommand{\fig}[1]{Fig.~\ref{#1}} 
\newcommand{\Fig}[1]{Figure~\ref{#1}}

\newcommand{\sect}[1]{Sec.~\ref{#1}}

\newcommand{\los}{${\rm LOS}$}

\newcommand{\NW}{network}
\newcommand{\IN}{internetwork}
\newcommand{\inw}{\textsl{INw}}
\newcommand{\nw}{\textsl{Nw}}
\newcommand{\brms}{$B_{\rm rms}$}
\newcommand{\fff}{$f$-mode\;}
\newcommand{\fffns}{$f$-mode}
\newcommand{\ef}{$E_f$}
\newcommand{\eft}{$\tilde{E_f}$}
\newcommand{\mef}{E_f}

\newcommand{\colfig}[3][1.]{\begin{figure}\centering
    \includegraphics[width=#1\linewidth,clip=TRUE]{#2}
    \caption{#3}
    \label{#2}
\end{figure}}
\newcommand{\colfigtwocol}[3][1.]{\begin{figure*}\centering
    \includegraphics[width=#1\linewidth,clip=TRUE]{#2}
    \caption{#3}
    \label{#2}
\end{figure*}}

\graphicspath{{./figures/}}

\begin{document}

\title{Solar-Cycle Variation of quiet-Sun Magnetism and Surface Gravity Oscillation Mode}
\titlerunning{Quiet-Sun Solar Variability}

\authorrunning{Korpi-Lagg et al.}

\author{M. J. Korpi-Lagg\inst{1,3} \and A. Korpi-Lagg\inst{1,2} \and N. Olspert\inst{1} \and H.-L. Truong\inst{1}}
\institute{Department of Computer Science, Aalto University, PO Box 15400, FI-00076 Aalto, Finland \and
Max Planck Institute for Solar System Research, Justus-von-Liebig-Weg 3, D-37077 G\"ottingen, Germany \and
Nordita, KTH Royal Institute of Technology and Stockholm University, 
              Hannes Alfv\'ens v\"ag 12, SE-10691 Stockholm, Sweden
}

\abstract{The origin of the quiet Sun magnetism is under debate. Investigating the solar cycle variation observationally in more detail can give us clues about how to resolve the controversies.}
{We investigate the solar cycle variation of the most magnetically quiet regions and
their surface gravity oscillation ($f$-) mode integrated 
energy $E_f$.
}
{We use 12 years of HMI data and apply a stringent selection criteria, based on
spatial and temporal quietness, to avoid any influence of active regions (ARs).
We develop an automated high-throughput pipeline to go through all available magnetogram data
and to compute 
$E_f$
for the selected quiet regions.}
{We observe a clear solar cycle dependence of the magnetic field strength in the most
quiet regions containing several supergranular cells. For patch sizes smaller than a supergranular
cell, no significant cycle dependence is detected. The $E_f$ at the supergranular scale 
is not constant over time.
During the late ascending phase of Cycle 24 (SC24, 2011-2012), it is roughly constant, but starts diminishing in 2013, as the maximum of SC24 is approached. This trend continues
until 
mid-2017,
when
hints of strengthening at higher southern latitudes
are seen.
Slow strengthening continues, stronger at higher latitudes than at the equatorial regions, but $E_f$ never returns back to the values seen in 2011-2012. Also, the strengthening trend continues past the solar minimum, to the years when SC25 is already clearly ascending. Hence $E_f$ behavior is not in phase with the solar cycle.
}
{The solar cycle dependence at the supergranular scale is indicative for the fluctuating magnetic
field being replenished by tangling from the large-scale magnetic field, and not solely due to
the action of a fluctuation dynamo process in the surface regions. 
The absence of variation at smaller scales might be an effect of the
limited spatial resolution and magnetic sensitivity of HMI.
The 
anticorrelation of $E_f$ with the solar cycle in gross terms is expected, but the phase shift of several years indicates a connection to the large-scale poloidal magnetic field component rather than the toroidal one. Calibrating AR signals with the QS $E_f$ does not reveal significant enhancement of the $f$-mode prior to AR emergence.
}


\maketitle

\section{Introduction} \label{sec:intro}

Localized regions of intense bipolar magnetic structures, called active regions
(ARs), are seen on the solar surface. Their numbers vary periodically 
in an 11-year cycle
and trace the
butterfly 
diagram,
which 
reveals 
a cyclic magnetic activity of the Sun in a latitude--time
domain. Such diagrams have proven to be useful and reveal some properties of 
the solar large-scale magnetic field. The large-scale magnetic field reverses its polarity every
second such cycle, which constitutes the solar magnetic cycle of 22 years. The origin of this
global variability is not yet fully understood \citep[see, e.g.,][]{char10}.

In addition to the global solar magnetic field, it is observationally known that there are ubiquitous small-scale magnetic fluctuations, the origin of which is equally debated - they could arise through the action of a small-scale dynamo instability or tangling of the large-scale magnetic field due to turbulence driven by convection, but it is not self-evident how the former mechanism could work in the solar plasma, where the conditions are in general unfavourable for it \citep[see, e.g.][]{BS05}. These magnetic fluctuations are usually studied by investigating the magnetically quiet regions on the solar surface (hereafter referred to as the quiet Sun, QS). Several observational
studies of QS magnetism have been conducted, and it has been argued
that the QS magnetic field is 
independent of the solar cycle 
\cite[see, e.g.,][]{Kleint+10,Buehler+13,Faurobert+15,Jin+15b,Jin+15a}. On the other hand, some other studies have proposed
that some dependence should exist \cite[e.g.][]{Lites+14,Meunier06,Faurobert+21}. 
Taken the theoretical and observational controversies, further studies of QS magnetism are required, and long-term investigations are now enabled by instruments like
Helioseismic and Magnetic Imager \cite[HMI,][]{2012SoPh..275..207S,2012SoPh..275..229S} on board the Solar Dynamics Observatory \cite[SDO,][]{2012SoPh..275....3P}
This instrument
provides high sensitivity, high spatial
resolution, long-term stability, and constant conditions. Currently, the data covers 12 years of observations, enabling solar-cycle-scale studies.

Another motivation to study the QS comes from recent observational investigations \cite[]{SRB16, Waidele22} that have reported 
 strengthening of solar surface or the fundamental \fff
about one to two days before
the formation of ARs using different kinds of local helioseismic techniques. Accompanied with numerical simulations that have given similar indications \cite{S+14,S+15,S+20}, this seems as a very promising avenue of studying the origin of solar sub-surface magnetism and the mechanism of active region formation. The observational studies, however, suffer from a lack of proper calibration method against the QS. \cite{SRB16}, for example, rely on QS
regions on the opposite hemisphere to compare with the AR \fffns. This method
requires that a quiet patch exists on the other hemisphere, and hence 
limits the number of ARs that can be included in the hindcasting procedure. It is
also prone to be affected by the probable fluctuations in the QS
\fff level. Although the results look promising, proper calibration with
statistically sound QS level, not just a comparison of a random QS patch on the
other hemisphere, is necessary to prove the robustness of these findings. Also,
such a calibration procedure is necessary for increasing the sample size. 
Building such a QS calibration data product is one of the main aims of this
study: we carefully identify the quietest regions on the solar surface based on the level of magnetic activity observed in
line-of-sight (LOS) magnetograms that are readily available from HMI, and
compute the \fff 
energy
at the central meridian as function of latitude
and time with suitable averaging. Building such a data product for the \fffns, requiring us to identify the most
inactive regions on the solar surface, allows us to extract statistics
of the QS magnetism as well, which is the second main aim of this study.

The paper is organized as follows: in Sect.~\ref{pipeline} we describe
the data, the necessary steps to 
clean it, and the automated pipeline we built for harvesting the data and compiling the end products, namely the QS magnetism data products, and the QS and AR \fff data. In
Sect.~\ref{results} we discuss our findings for the QS magnetism, \fffns, and the lastly present some AR data with the QS calibration
applied.

\section{Observations}\label{pipeline}

Our analysis is based on data from HMI@SDO. We use two standard data products: (i) full-disk line-of-sight (LOS) magnetograms, computed every 720\,s by combining filtergrams obtained over a time interval of 1260\,s (\texttt{hmi.M\_720s}), and (ii) full-disk LOS dopplergrams, computed every 45\,s from six positions across the nominal 6173.3\,\AA{} spectral line (\texttt{hmi.V\_45s}). We processed the two data sets in a semi-automatic pipeline (see \fig{pipeline}), optimized for obtaining reliable information about the magnetic field in the QS regions and for a robust computation of the \fff power from the dopplergrams. The left tree in \fig{pipeline} describes the pipeline used for the magnetograms, the right tree for the dopplergrams.
In the following we use the following notation and definitions. 
We denote the solar 
latitude with $\lambda$, longitude with $\varphi$, 
both in the Stonyhurst coordinate system \cite[]{Thomson06}, 
and time with $t$.

\colfig{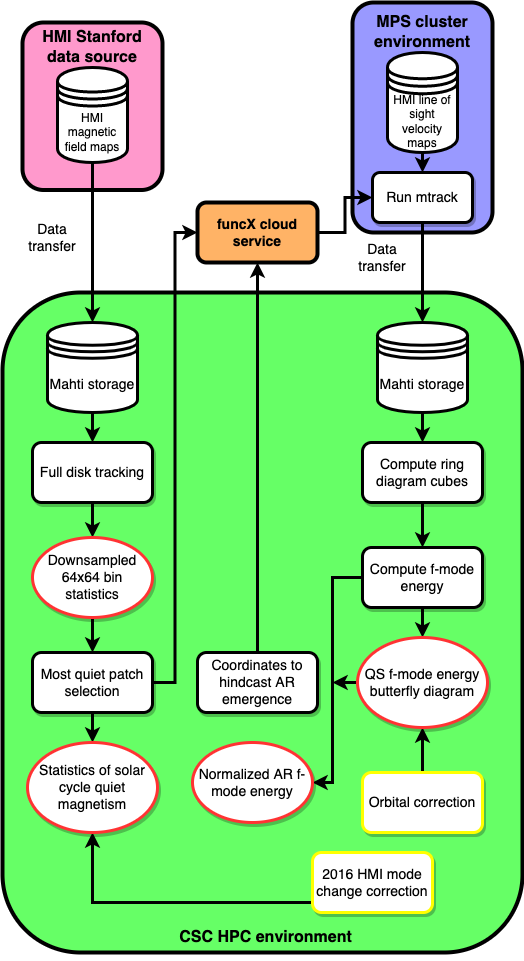}{Data pipeline. The leftmost 
tree illustrates the pipeline to collect the magnetograms in the most quiet patches, while the rightmost one the pipeline to track data from the quiet patches from the HMI database, and compute the \fff energy. The central path of AR processing is otherwise equivalent to the QS pipeline, but there AR coordinates are sent for tracking in the 
MPS (in Germany) cluster environment, and an additional step, namely QS calibration, is performed in the end. Rectangular boxes represent analysis functions, and ellipsoids the derived data products. The rectangles with yellow frames stand for the corrective functions applied to the data. }

\subsection{Magnetograms}

The first data product, full-disk line-of-sight (LOS) magnetograms, provides a direct measurement of the variability of the QS
magnetism during a solar cycle. To enhance the signal-to-noise ratio in the LOS magnetograms we performed a newly developed algorithm for spatial and temporal averaging: The full-disk HMI magnetograms starting from 27-Apr-2010 and ending at 
04-May-2022
were downloaded from the Joint Science Operation Center (JSOC) hosted at Stanford University (http://jsoc.stanford.edu) to a temporary storage (see \fig{pipeline}, 'Mahti storage') and tracked at full spatial resolution for 8 hours to compensate for the solar rotation ('Full disk tracking').
We neglected differential rotation due to
the short tracking time.
The step between tracked sequences was 4 hours, so that a total of 6 tracked sequences were gathered per one day, resulting in more than 
$25765$ tracked sequences (as of 04-May-2022).

From each tracked sequence, we extracted two data products by dividing the visible solar disk  between latitudes and longitudes from $-80\degree$ to +$80\degree$ into (i) $64\times 64$  overlapping patches of $15\degree$ (in solar latitude and longitude)  and (ii) $180\times 180$ patches of $1\degree$.
Every of these patches therefore contains a space-time cube of LOS magnetograms at full spatial resolution at a 12-minute cadence,
from which we 
compute the 
root-mean-square
(rms) 
of the magnetic field strength, averaged over the full field-of-view of the tracked cube 
and over the full 8 hour period
(\brms{}=$\sqrt{\langle B^2\rangle_{\rm LOS}}$).
The $15\degree$ patches (i) are large enough to cover several supergranulation cells containing \NW{} and \IN{} fields \cite[]{2010LRSP....7....2R} with a typical size of 30--35\,Mm (we refer to them as \nw{} cubes), and the $1\degree$ patches (ii) are small enough that some of them lie completely in the \IN{} (\inw{} cubes). 
The statistics for each patch were stored as data products (see also \fig{pipeline}) including the information about 
latitudinal and longitudinal position as well as the Carrington 
longitude for \NW{} and \IN{}. We refer to these maps as the \nw{} and the \inw{} statistical maps.

\subsection{Quiet region selection\label{quietregion}}

From the 
statistics computed from the cubes, 
\brms{}
turned out to be the best tracer for determining the magnetic activity level. It could clearly distinguish between $15\degree$ patches containing active regions, plage, enhanced network and quiet network. Also, it depicted very well the low-field \IN{} regions.

The analysis of the solar-cycle variation of the 
QS
magnetism required a careful selection of the most quiet regions, defined as being free of enhanced solar activity. We therefore searched for the minimum value of \brms{} in both, the \nw{} and the \inw{} statistical maps, on a latitudinal grid with a $10\degree$ spacing fulfilling the following additional criteria:
\begin{enumerate}[(i)]
	\item\label{c1} the most quiet pixel must be within $\pm10\degree$ around the central meridian,
	\item\label{c2}  this pixel must belong to the 10\% most quiet pixels of the month,
	\item\label{c3}  this pixel must be the most quiet pixel within a 4-day interval.
\end{enumerate}

Criterion (\ref{c1}) was chosen to get the strongest magnetic field signal along the central meridian. 
Criterion
(\ref{c2}) guarantees an equal distribution of quiet pixels over the 12-year period of available HMI measurements, and 
criterion
(\ref{c3}) ensures that the quiet pixels for the 1-month period do not originate from the same supergranular structure, since the dynamical evolution time of the supergranulation lies between 24 and 48\,h \cite[]{2010LRSP....7....2R}. 
As a result of applying these criteria we obtained
two time series of the \brms{} for the most quiet patches in the \NW{} and the \IN{} regions. 
Note that this selection also efficiently removes the 24\,h modulation present in the HMI magnetograms.

\subsection{Correction for HMI sensitivity change\label{sensicorr}}

The temporal evolution of \inw{} \brms{} value
revealed clearly a change in the HMI observing mode, performed on 13-Apr-2016. On this day, HMI switched to a more efficient observing mode \cite[see][]{2018SoPh..293...45H,2014SoPh..289.3483H,2016SoPh..291.1887C}. By combining both HMI cameras to determine the vector-field observables the cadence for full-disk magnetograms could be reduced from 135\,s (observational mode MOD-C) to 90\,s (MOD-L). This reduced the noise level for Stokes $V$ measurements by 17\%, resulting in a decrease of the noise level in the \los{} magnetograms by 5\%.

For the long-term study presented in this paper, we need to correct for this sensitivity change. A very accurate correction method can be derived from the \inw{} time series: since it contains only the most quiet pixels over a certain latitude region and time, the sensitivity change results in a step function. The value of the step was determined by fitting a polynomial to the \brms{} values determined from the \inw{} time series plus a Heaviside step function, centered at the date of the mode change. We used the Bayesian information criterion \cite[BIC,][]{Stoica2004} to determine the degree of the polynomial, which lies between 1 and 6 for the various latitudes. We want to note that the retrieved amplitude of the Heaviside step function is only weakly dependent on the degree of the polynomial. This fitting is exemplified in \fig{fit-offset} for the solar latitude 0$^\circ$, where the minimum value for the BIC was reached for a fit with a polynomial of 
degree 3. 
The so-determined amplitude of the Heaviside step function is added to the 
\brms{}
data points after 13-Apr-2016 for all data presented in this paper.

\colfig{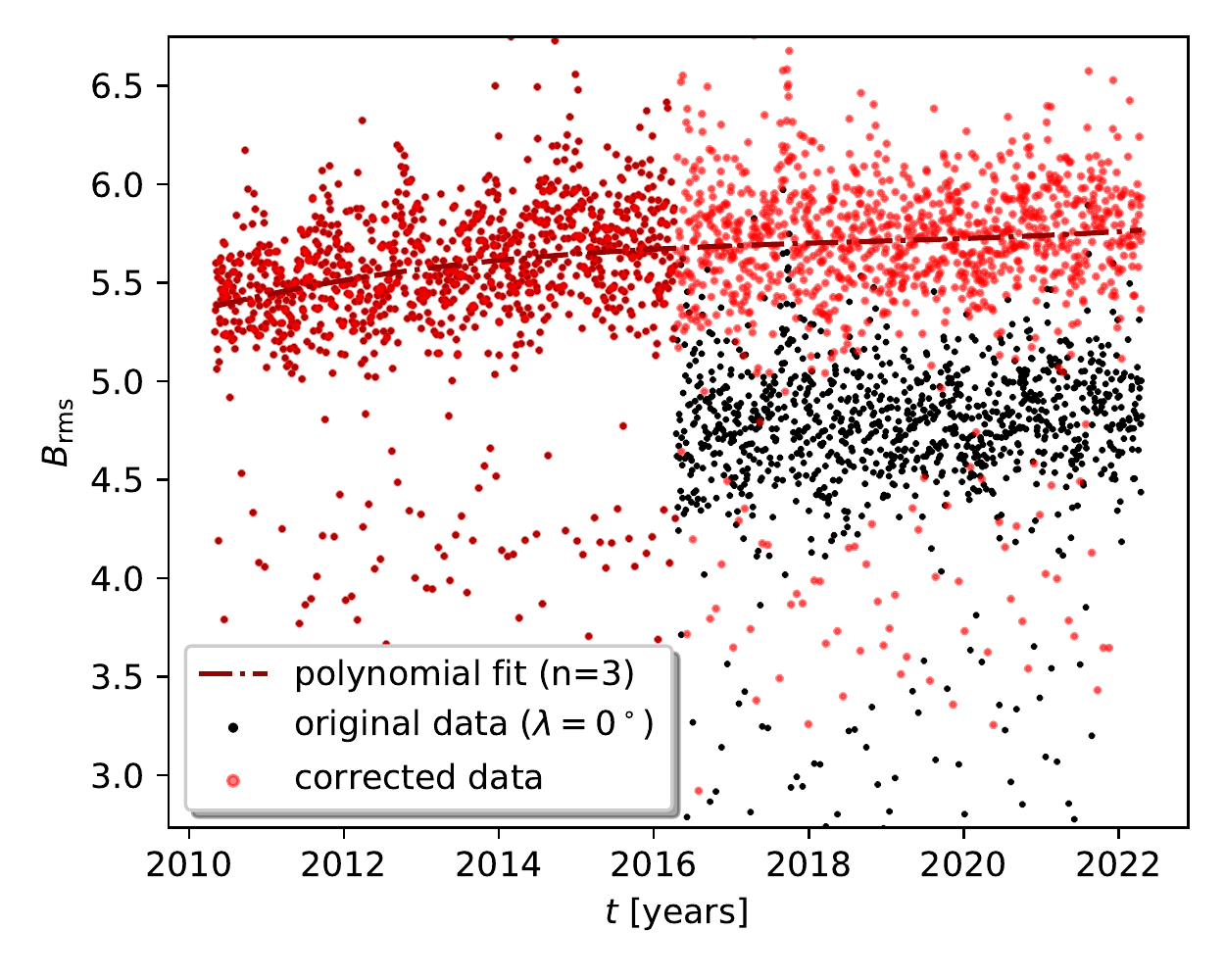}{Determination of the correction for the HMI sensitivity change: the observing mode change on April 13 2016 causes a discontinuity in the level of \brms{} values of the \IN{} data. The original data are displayed in with the dark red and black dots, the corrected data with the light red dots. The dashed line indicates the polynomial fit of degree 3 used to obtain the offset.}

\subsection{Dopplergrams}

The second HMI data product used in this paper are the \los{} velocity maps. The goal is to compute the 
energy contained in the
surface gravity mode, the so-called \fffns,
and investigate whether variability during the solar cycle is present. Such variations could be caused by the presence of non-emerging sub-surface magnetic fields, as the \fff is known
to be strongly affected by the presence of magnetic fields \citep[see, e.g.,][]{Cally+94,CB97,SRB16}. 
It is also very important to study this question in more detail, for establishing a reliable calibration method 
for a measurement of
the claimed 
\fff enhancement prior to AR emergence \cite{SRB16,Waidele22}. 

The dopplergram data is hosted in the German Data Center for SDO (GDC-SDO) on a server at the Max Planck Institute for Solar System Research (MPS Göttingen, Germany) whereas the analysis is executed in the CSC supercomputing environment (Finland). We have utilized  a function-as-a-service client based on funcX \cite[]{chard20funcx} for accessing the required data in the database server. We have developed functions based on funcX API and deployed the functions in the MPS environment. These functions leverage the mtrack\footnote{\url{http://hmi.stanford.edu/teams/rings/modules/mtrack/v25.html}} command to prepare the dopplergram cubes within the GDC-SDO environment and subsequently transfer them to the CSC environment. The coordinates of the selected 
QS
regions were sent to the funcX service, which invokes suitable functions to automate the data retrieval and movement.
The data processing pipeline used Astropy\footnote{\url{http://www.astropy.org}}, a community-developed core Python package for Astronomy \citep{astropy13, astropy18} and version 3.1.6 of the SunPy\footnote{\url{https://sunpy.org}} open source software package \cite[]{sunpy20}.

\subsection{Computation of the \fff
energy}

\begin{figure}\centering
	\includegraphics[width=1.0\linewidth]{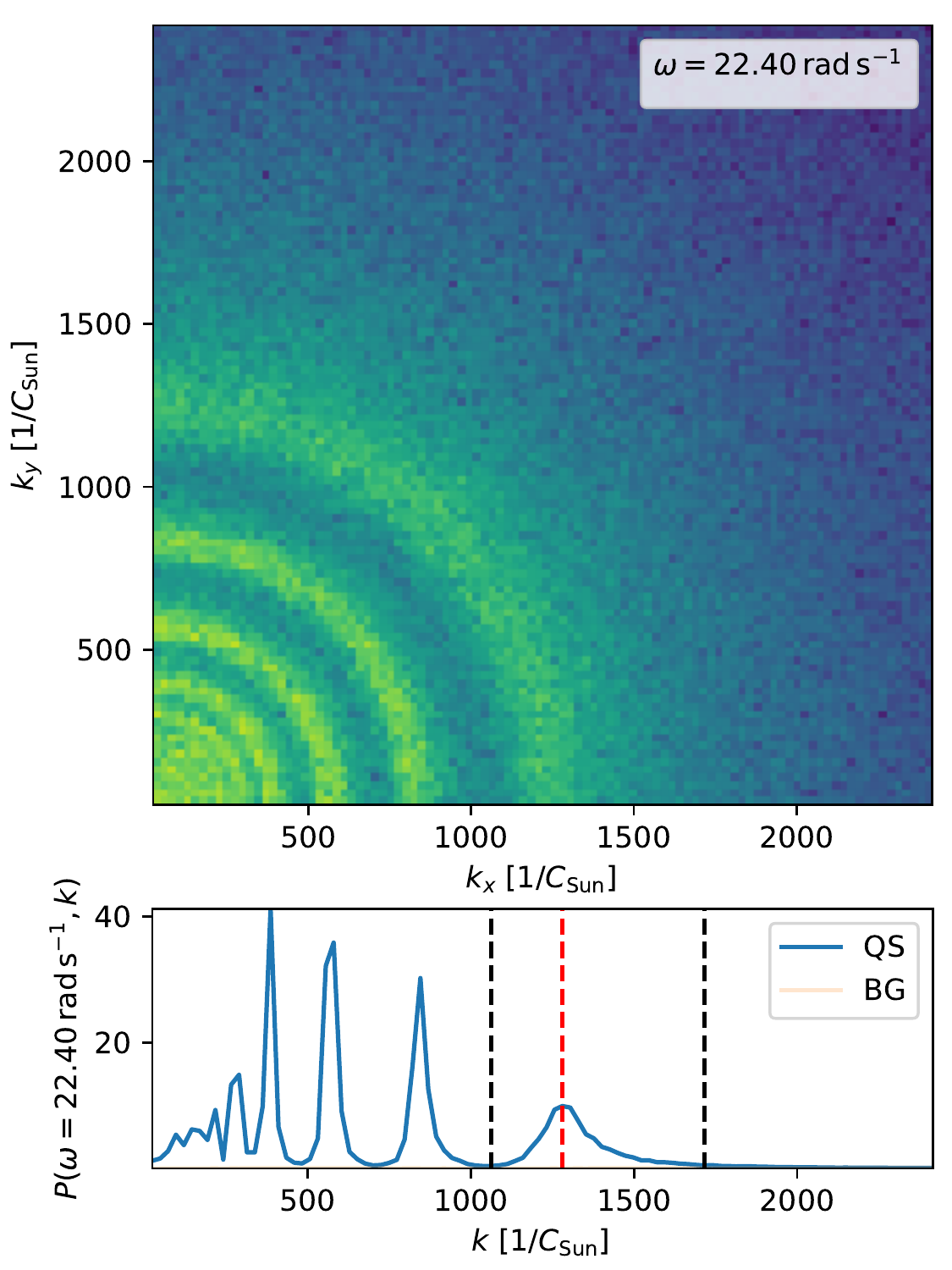}
	\caption{Top: example of quiet-sun ring diagram at
	the angular frequency 
	$\omega=22.40$\,rad\,s$^{-1}$ 
	for one quadrant of the full ring (logarithmic color scale), computed from the data cube presented in \fig{2010-07-21_2014-11-09} (left panel). 
	Bottom: collapsed spectrum obtained by 
	integrating over
	the ring diagram. The red vertical line marks the position of the maximum of the \fff and the black vertical lines the range of integration. The wavenumber $k$ is expressed in units of the circumference of the Sun ($C_{\rm Sun}$).}
	\label{ring_diagram}
\end{figure}

Subsequent processing involved calculating the 
three-dimensional
power spectra for each dopplergram cube and 
integrating
in $(k_x, k_y)$-plane, for each 
angular
frequency $\omega$,
over all wavenumbers, 
$k=\sqrt{k_x^2+k_y^2}$,
to obtain a collapsed power spectrum, $P(\omega,k)$. An illustrative case for a frequency
value at which \fff is strong, is shown in 
\fig{ring_diagram},
upper panel.
Such a procedure
significantly reduced the noise level leading to smooth one-dimensional $k-\omega$ spectra,
an example being shown in the lower panel of \fig{ring_diagram};
in this kind of a plot, the \fff is the rightmost peak at the highest $k$ values.
The adopted procedure
is justified for the 
QS
spectra, as the ring diagrams are radially symmetrical w.r.t. $\omega$ axis. 
To obtain the total energy contained in the \fffns, $E_f$, we
performed another integration over the separated \fff signal.
In contrast to earlier studies \citep{SRB16,Waidele22}, we
did not perform fitting to the \fff for its extraction. 
Instead, we only determined the background signal, and subtracted this contribution.
The background determination used the fact that the position of the \fff peak for low values for $\omega$ (between 3 and 6 rad\,s$^{-1}$) lies outside the $k$ range used for the computation of the \fff power. Since the background level does not change significantly with $\omega$, the average spectrum over this $\omega$ range provided a good estimate for the background signal.
Then,
an integration range in $k$ space was selected for each constant $\omega$ in a following way: first the maximum of \fff $k_{\rm max}$ and the minimum between \fff and first $p$-mode $k_{\rm start}$ were detected; $k_{\rm start}$ was chosen as the start of integration range and the end of the integration range $k_{\rm end}$ was set as $k_{\rm end}=k_{\rm max}+2(k_{\rm max}-k_{\rm start})$; $k_{\rm end}$ chosen in such a way guarantees that the integration range is sufficiently wide to cover significant part of the \fff 
signal in $k$-space.
Integration range in $\omega$ space was chosen between values starting from 
14.45 and ending at 29.03, 
where the significant part of the \fff power resides. 
As the end result, we have computed the \fff energy
\begin{equation}
E_f=
\sum_{\mbox{$k_{\rm start}$}}^{\mbox{$k_{\rm end}$}}
\sum_{\mbox{$\omega_{\rm min}$}}^{\mbox{$\omega_{\rm max}$}} P(\omega,k)
\end{equation}
After completing the described steps in the CSC supercomputing
environment, we have collected a total of 
22680
\fff
energies
for patches close to central meridian over all latitudes covering 
most of SC24 and the ascending phase of SC25.

\subsubsection*{Orbital correction of the \fff{} 
energy}

Since the \fff
energy
is computed from the \los{} dopplergrams, its value depends strongly on the viewing geometry. This dependence,  roughly following the cosine of the solar latitude for data taken at the central meridian, is additionally modulated by the orbital motion of the Earth around the Sun, which changes the viewing angle at any given solar latitude by $\approx\pm7^\circ$ during one year. We compensated for this periodic variation by fitting the parameters ($x_0(\lambda), ..., x_3(\lambda)$) of the following function to all observations of a given latitude $\lambda$:
\begin{eqnarray}
\label{eq:orbitcorr}
A_{\mbox{corr}}(\lambda) &=& x_0(\lambda) (  2 ( (1+\cos(\Phi+x_1(\lambda)))/2)^{x_3(\lambda)}-1   )\nonumber \\
&+&x_2(\lambda),
\end{eqnarray}
with $\Phi$ being the phase angle of the Earth defined as the cotangens computed from the $x,y$ barycentric position of the Earth.
This correction $A_{\mbox{corr}}(\lambda)$ is then subtracted from the computed 
$E_f$ values, which efficiently removes any yearly variation. 

\section{Results}\label{results}

After applying the selection criteria described in \sect{quietregion} and the subsequent correction for the 
HMI's
sensitivity change (\sect{sensicorr}) we obtain the dependency of the \brms{} values of the most quiet region at the central longitude as a function of heliographic latitude and time. \fig{Brms-lat00-percentile} presents the \brms{} values from May 2010 until May 2022 for the 
Stonyhurst latitude and longitude $(\lambda,\varphi) = (0,0)^\circ$.
The individual data points (red dots) are computed over an area of 15$^\circ$ in latitude and longitude and over a time of 8 hours. At disk center, this corresponds to an area of $\approx 180 \times 180$\,Mm$^2$, and therefore contains 30--40 supergranular cells. The \brms{} value therefore contains
\NW{} and \IN{} fields.
Typical magnetograms at disk center at solar minimum (July 2010) and maximum (Nov 2014) are presented in \fig{2010-07-21_2014-11-09}. There is no obvious difference between the two maps: both show similar minimum and maximum field strengths, and the visible impression of the supergranular cells with the strong field patches of both polarities surrounding the \IN{} regions is indistinguishable.

\fig{Brms-lat00-percentile} clearly indicates the solar cycle dependence of the 
QS
magnetic fields, consisting of \NW{} and \IN{}. The \brms{} values peak in the second half of 2014, 
around half a year
later compared to the
declared maximum of solar cycle \#24 (April 2014, source: WDC-SILSO, Royal Observatory of Belgium, Brussels). The variation is statistically significant, as indicated by the the mode and the 97\% percentiles, computed by fitting a log-normal distribution over a 1-year sliding window.

A similar plot is presented in \fig{Brms-lat00-percentile-IN}, but now the \brms{} value was computed only for a $1\degree$ window in latitude and longitude, corresponding to an area of only $12\times 12$\,Mm$^2$ at disk center. The quiet-region selection criteria (\sect{quietregion}) guarantees, that this patch lies fully within the \IN{}, and is not `contaminated' by \NW{} fields. Clearly, there is no dependence of \brms{} with solar cycle detectable. The 
slight increase over the 12-year period might be an effect of the aging of the HMI instrument and the resulting decrease in the sensitivity for magnetic field measurements.

The average \brms{} value in the \IN{} is in $\approx6$\,G. Using the area of one HMI pixel this corresponds to a magnetic flux of $3\times 10^{16}$\,Mx. 
This value is certainly affected by the sensitivity of the HMI instrument, and is reduced to $\lesssim5$\,G after the HMI sensitivity change described in Sect.~\ref{sensicorr}.

\colfigtwocol{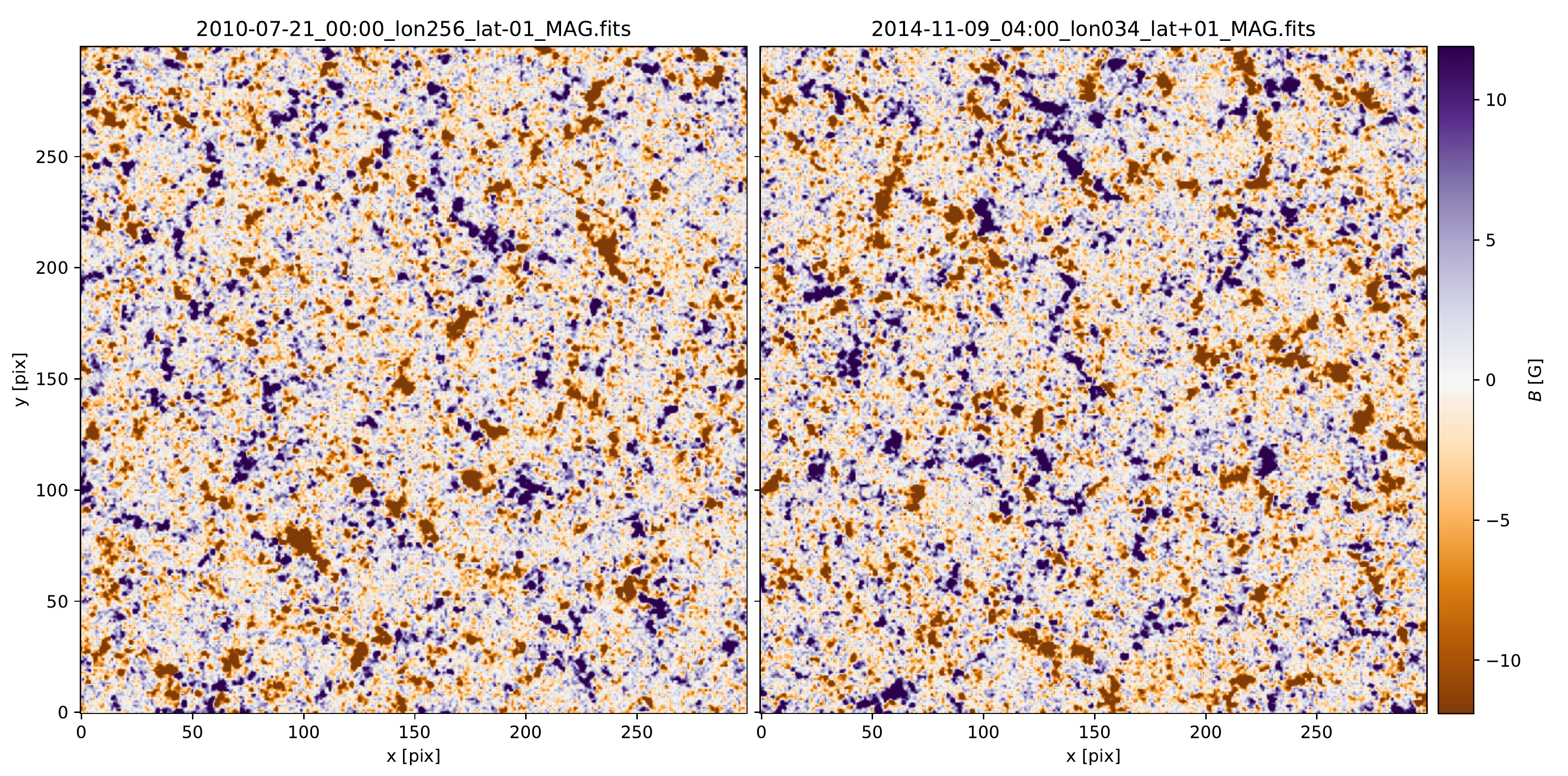}{Comparison of two disk-center magnetograms at solar minimum (2010-07-21) and 
around the maximum seen in the disk-center \brms{} values
(2014-11-09). The maps show the first frame of the 8\,h data cube.}

\colfig{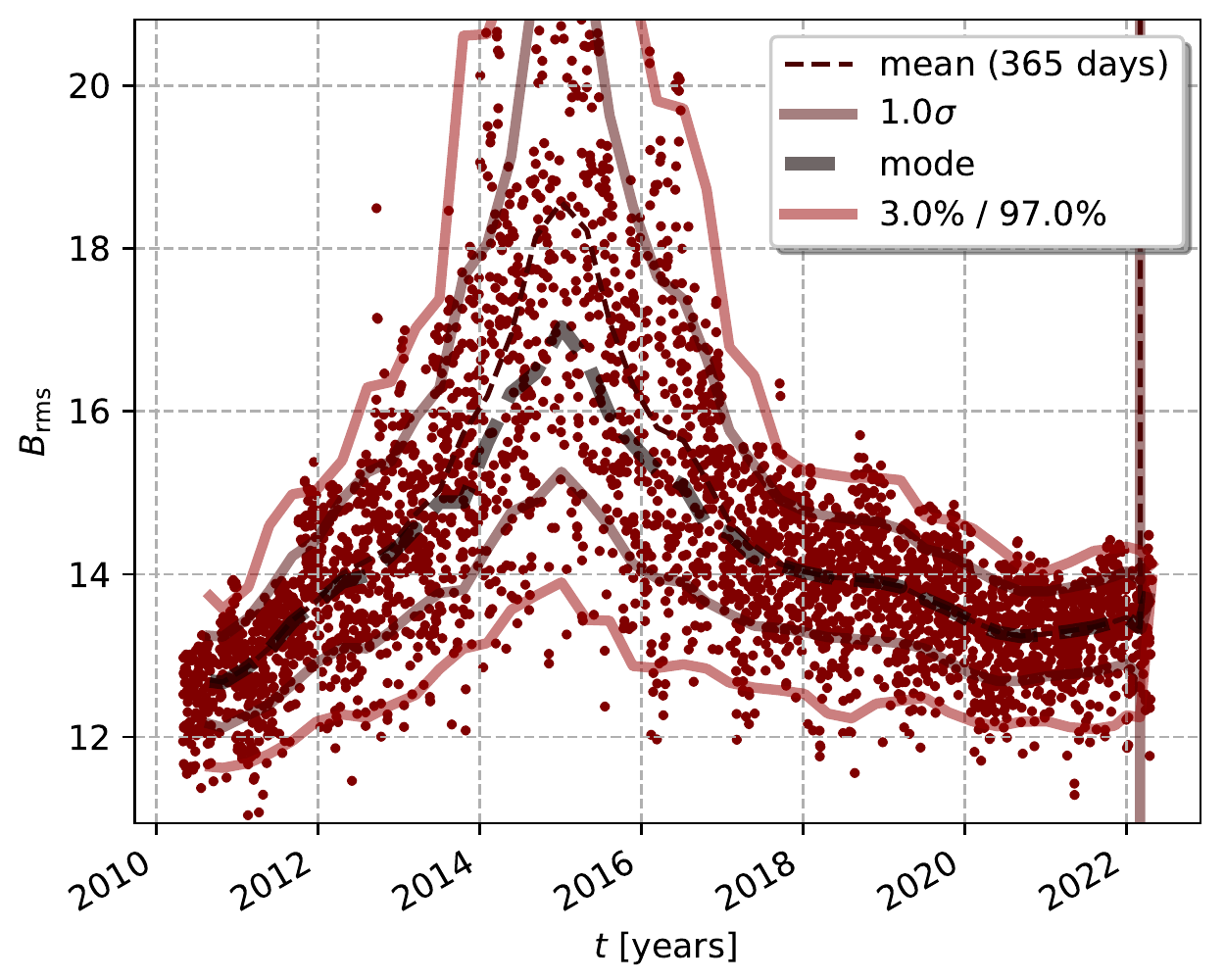}{\brms{} determined from the most quiet patches from 2010 to 
2022
at disk center. The patches of 15$^\circ$ size in longitude and latitude contain \NW{} and \IN{} fields. The solid red lines display the 97\% percentile level, the dashed, gray line the mode of a log-normal fit to yearly-binned data moved in a sliding window of 100 days length.
For all values of $\omega$ the magnitude of $P$ is always an order of magnitude higher than
the noise level outside the considered location of the ring.
}

\colfig{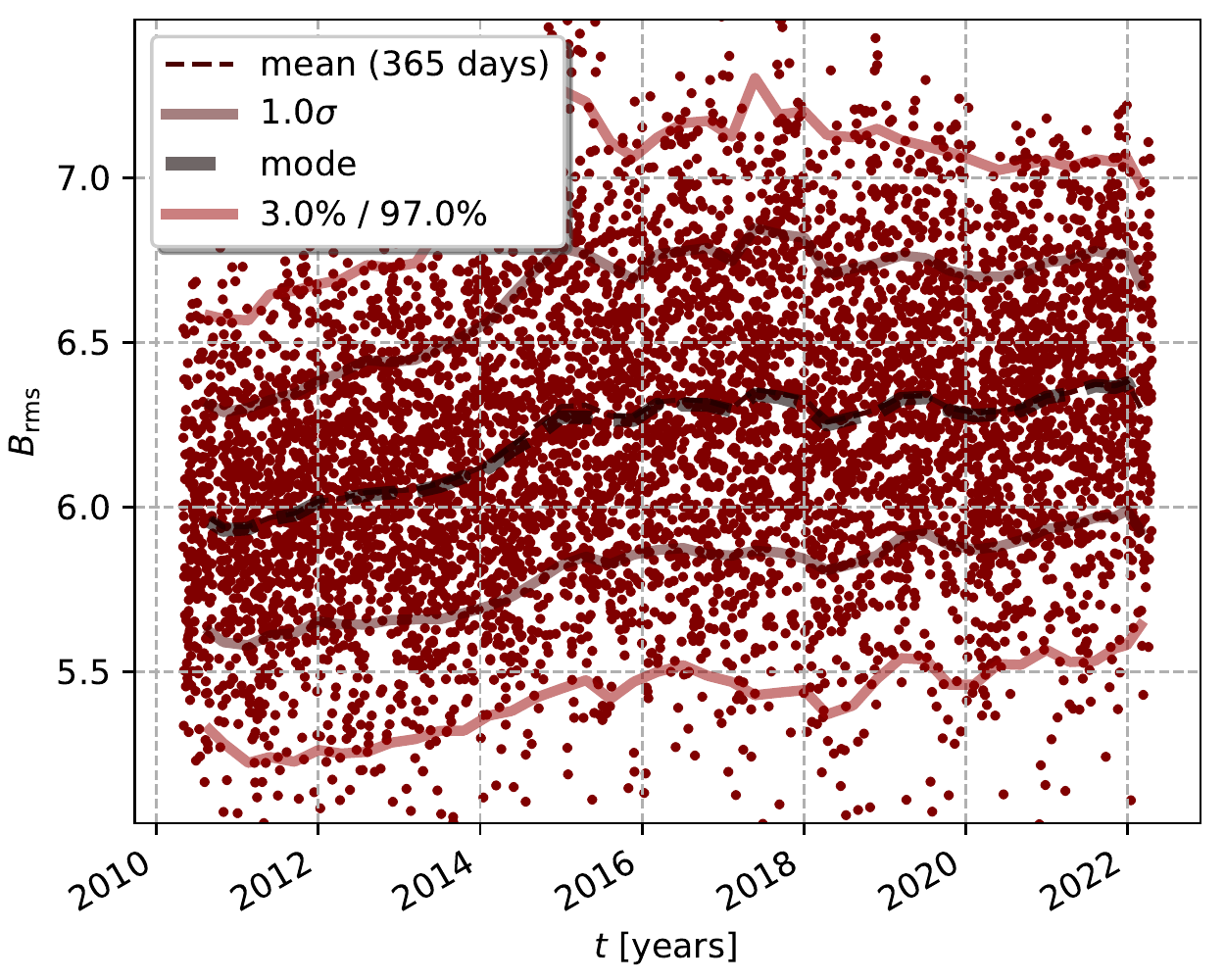}{Same as \fig{Brms-lat00-percentile}, but for \brms{} determined from patch size of only of 1$^\circ$ in longitude and latitude, corresponding to \IN{} patches. }

\subsection{Quiet Sun \fff butterfly diagram}\label{qsf}

To investigate whether the \fff energy has any significant solar cycle dependence, we construct butterfly-diagram-like plots by computing and removing the temporal mean, and plotting the so-obtained data in a binned latitude-time diagram. 
We compute the temporal mean over the whole time series as 
\begin{equation}
    \overline{\mef}(\lambda) = 1/n_t \sum_t \mef(\lambda,t),
\end{equation}
where $n_t$ is the total number of time points at a certain $\lambda$, and denote it with an overbar. Next, we compute the variation of the \fff energy around this level, defined as
\begin{equation}
    \mef'(\lambda,t)=\mef(\lambda,t)-\overline{\mef}(\lambda),
\end{equation}
and plot this quantity
for the central meridian
as function of latitude and time as in \Fig{butterfly_lat}, 
in patches 
with a 3-month binning in time and roughly 6 degrees in latitude. 

To estimate 
the overall variability level of the QS, and compare it to the cycle variation level defined above,
we will use the standard deviation of the data in a patch.
We take $N_t$ bins in time, $N_\lambda$ in latitude, and denote the number of \ef measurements contained in such a patch by $N_P$. We compute the average \fff energy in a patch as
\begin{equation}
\langle \mef \rangle=1/N_P \sum_{N_P} \mef (\lambda,t).
\end{equation}
Then we can straightforwardly calculate the standard deviation as
\begin{equation}
\sigma= \sqrt{\frac{1}{N_P}\sum_{\mbox{patch}} ( \mef(\lambda,t)
- \left<{\mef} \right>)^2}.
\end{equation}
From this we form the average error at each latitude by taking an average over time and denote it as $\overline{\sigma}$. 
This quantity is plotted as the error bar in the right panel of 
\fig{butterfly_lat}.
By inspecting the level of variability around the temporal mean and the 
overall variability of the QS, we must conclude that these signals are of the same order of magnitude. 
In this respect, the solar cycle dependent signal is very weak, but if we compare that to the noise level in the data, both the QS variability level and cycle-dependent signal exceed the noise level by an order of magnitude, hence this is not an issue concerned with the instrument sensitivity.

What could be the cause of the strong QS variability level is most likely related to the strongly varying sub-surface magnetic fields that affect the \fffns,
invisible in
the surface \brms{} which was used as the selection criterion for really quiet patches. 
This is illustrated in \fig{fmode_vs_brms}, where we show the measured \fff energy versus the \brms{} in the patches. The \fff energy is not correlated with the \brms{} seen at the surface, and strong, nearly constant, variability is seen at different epochs of the solar cycle. This poses clear limitations of the calibration method proposed here to detect weak transient signals in the \fff evolution.

\colfigtwocol{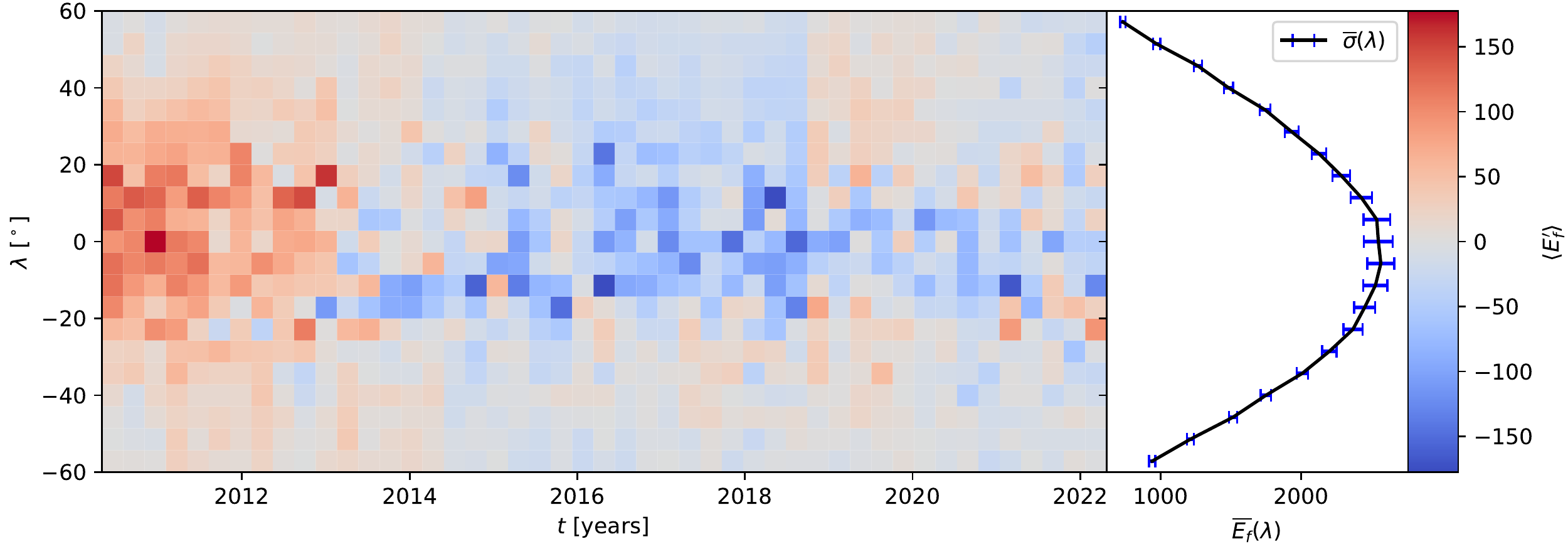}{Butterfly diagram of \fff  
energy variations, $E'_f(\lambda,t)$, \todo{we should still describe the binning}.
computed from the collapsed ring diagrams (see bottom panel of \fig{ring_diagram}).
The temporal average, $\overline{\mef}(\lambda)$, 
has been
subtracted from the data, 
and
is shown as the black line in the right panel. 
Its error bars
represent the time-averaged standard deviation of the fluctuations of the \fff 
energy
around the signal for every bin ($\overline{\sigma} (\lambda)$). 
}

\colfig{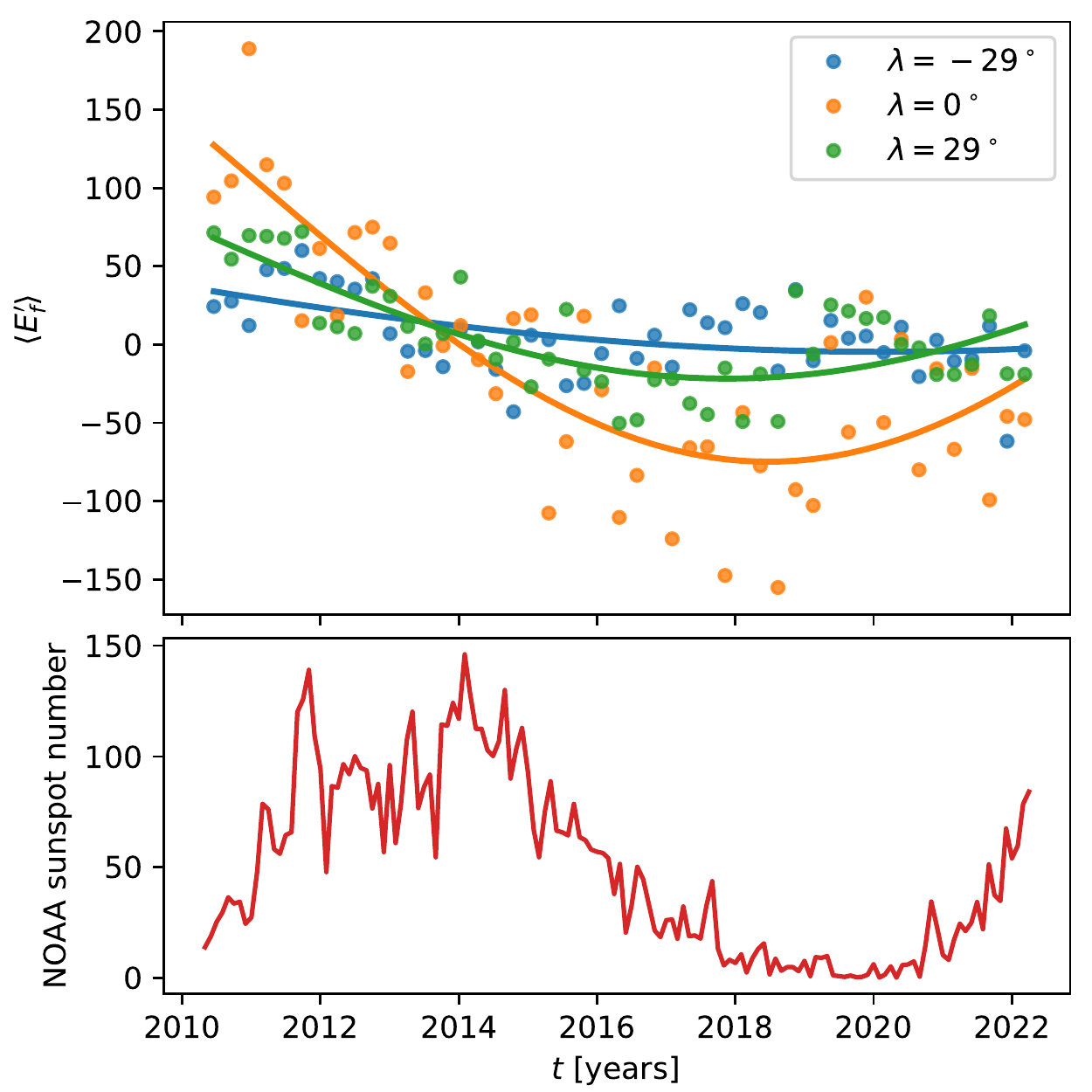}{Top: Cut through butterfly diagram of \fff energy variation for three different latitudes (see legend). The solid lines represent a fit using a sine function. Bottom: NOAA sunspot number.}

\Fig{butterfly_lat} displays the butterfly diagram of the QS
\fff energy determined
from $15\degree \times 15\degree$ patches that showed only weak, small-scale magnetic structures.
To compute the \fff
energy, we have used the whole ring diagram data.
Temporally averaged profile in latitude is removed for highlighting the variation in the
\fff power. The period covered is somewhat larger than the 11 year 
SC24,
which
began to show some magnetic activity from early 2010; this cycle ended by around 2020 (see also the lower panel of \Fig{butterfly_lat_cut}, showing the sunspot number of SC24 and SC25 up to the writing of this manuscript).
Broadly, we find an anti-correlation of the \fff 
energy with the activity cycle of the
Sun --- during the solar minimum and early rising phase ($\sim 2010-12$), the \fff
energies
are larger, and around cycle maximum to the declining phase,
the \fff
energy 
is suppressed.
This correlation is strongest near the solar equator, within about $\pm 20\degree$ in
latitude, and it decreases with higher latitudes.
Interestingly, we observe a temporal shift in the \fff butterfly diagram with respect to the
solar cycle, such that the minimum of the \fff occurs later, i.e., in the declining phase of
SC24, roughly around the end of the year 2016.

During mid-2017, we see weak strengthening of the \fff strength to commence, first occurring at higher Southern latitudes. During the minimum (2018-2019) between the SC24 and 25, the strengthening appears stronger in the higher latitudes. It also continues after the minimum, when 
SC25
has already started its ascending phase. What is also noteworthy is that the \fff
energy
never rises as high as the values observed during the ascending phase of SC24 (2011-2012).

In \Fig{butterfly_lat_cut} we present cuts through the butterfly
diagram at three different latitudes, equator, and $\pm 29\degree$, and sinusoidal fits to the data points. The variability is the strongest at the equator, but also contains the largest scatter. The minimum \fff energy 
is obtained there 
around the beginning of the year 2018. There, the difference in magnitude during SC24 and 25 is the largest. Southern higher latitudes show very weak variation in contrast to the equator, 
while the Northern one with an intermediate magnitude. 
The shift w.r.t. the solar cycle is somewhat latitude dependent. At the higher northern latitude, the \fff minimum occurs roughly a year earlier than at the equator. The shift in the \fff mode phase w.r.t. the sunspot cycle is of the order of three to four years, hence it is more in phase with the 
poloidal phase of the magnetic field than the toroidal one.

\subsection{AR \fff with QS calibration}\label{arf}

In this section we compare the \fff time evolution before and after the emergence of 
two ARs (11130 in \Fig{AR11130_track} and 11105 in \Fig{AR11105_track}), that were reported to show an enhancement both in \cite{SRB16} and \cite{Waidele22} w.r.t. a QS control patch
in the opposite hemisphere several days before their emergence. The main difference in our analysis
is that the \fff energy is now normalized to the actually measured average QS level
around the specific time and latitude. To minimize
the QS fluctuations, Gaussian smoothing to the QS \ef\,is applied, 
but the smoothing kernel width used is kept at 
$5^\circ$ in longitudinal and latitudinal direction
so that the fit to the data can be considered accurate.
\fff energy of 1 is hence
equivalent to that of the QS, and values stronger/weaker 
indicate enhancement/quenching of the \fff w.r.t. QS.
Another difference is that we use the
full ring diagram when computing the \fff energy, while the
other papers used only $k_x=0$ cuts. We have, however, performed
analysis with $k_x=0$ and $k_y=0$ cuts, and see no significant
difference between the two, nor w.r.t. the full ring data, except for the
increased noise level in \ef\,derived from single cuts.
As the measurement is done exactly in the same
manner for the QS and ARs, we anyway anticipate
that this difference should not influence the results significantly. 
Also, as more integration is performed, the less noisy is the 
signal, hence improving the data quality. 

The results from our AR analysis are shown with green symbols and lines in \Fig{AR11130_track} and \Fig{AR11105_track}: in the top panel, each orange point shows the normalized \fff energy, \eft, with four-hour cadence, 
smoothed with a 24-hour boxcar to remove a daily fluctuation
otherwise prominently present in the HMI data. The shaded areas show the QS variance, $E'_f$,
determined from the standard deviation of the variation presented in \fig{fmode_vs_brms},
is roughly constant (around 4--5 percent) at all latitudes and longitudes. When it is used as normalization of the \fffns, then the uncertainty at large longitudes becomes very high, as is indicated by the rapidly 
fanning 
shaded areas toward the limbs. The \brms{} measured simultaneously at the surface is shown in the lowest panels of the figures with black symbols. 

In the middle panels of \Fig{AR11130_track} and \Fig{AR11105_track}, we compare the different calibration curve candidates.
The calibration curve used by 
\cite{SRB16},
based on more simple cosinus dependence of the form $\cos{\alpha} \left[q + \left(1-q \right) \cos{\alpha} \right]$ with $q=0.5$
and $\cos\alpha = \cos\lambda \cos\varphi$, is shown with blue symbols and lines in the aforementioned panels.
As this specific curve does not fit the QS data very well, we also performed a fit and determined the optimal value of $q$, that turns out to be $-0.045$; this curve and calibration with it are plotted with orange symbols and lines in the top and middle panel of the figures.
The differences between these calibration curves are illustrated in the middle panels: The \cite{SRB16} calibration
curve does not fit the data, overplotted with gray dots, well at larger longitudes. There, it 
is decreasing less strongly than the data. Near the disk center the amplitude is too low to present the data well.
The new cosinus fit and the Gaussian convolution applied to the QS
data represent it better, with the consequence that the calibration curves decrease more steeply toward the limbs.
The fitted cosinus curve has somewhat larger amplitude than the Gaussian convolution. This has the effect that the \eft\,calibrated
with the cosinus fit tends to be always below the QS level. The Gaussian convolution calibration results at \eft\,values maximally at the QS level before the AR emergence.

As can be seen
from \Fig{AR11130_track} and \Fig{AR11105_track} top panels, 
applying the new QS calibration curve flattens the signal profile in comparison to the
earlier $q=0.5$ calibration used. For both of the selected ARs, the \fff energy close to the limb is below the QS level to start off with, which could indicate that both of these active regions occur in an environment
already affected by a sub-surface magnetic field. 
Applying the earlier calibration curve, a clear enhancement above the QS level is visible for AR11130, while for AR11105 the profile is similar, but the enhancement amplitude remains within the level of QS variability (shaded area). The new calibration shows a more modest enhancement, reaching its peak value at similar times as with the earlier calibration for AR11130, while somewhat earlier for AR11105, but its amplitude is not exceeding QS level significantly. Unfortunately, the possible enhancement of \fff energy of the selected ARs occurs so close to the limb that it is not completely certain whether it is an artefact of the rather uncertain limb calibration data, or a true effect. Obviously, a larger sample size should be studied, including also ARs that emerge closer to the west limb to completely prove/refute this scenario.  

\colfig{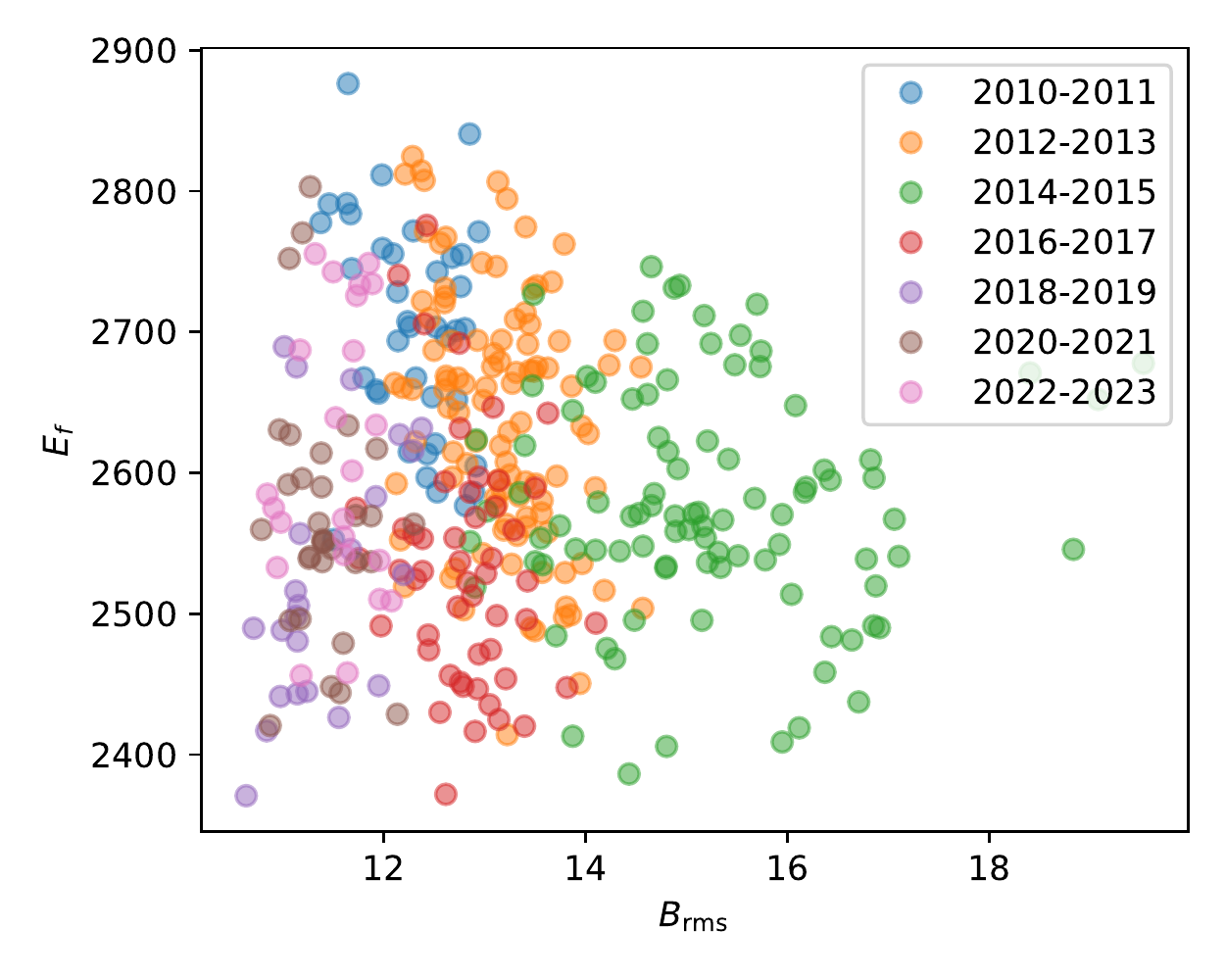}{\fff
energy
as a function of \brms{} for the most quiet patches at disk center. The colors indicate the grouping according to the years. The standard deviation for the \fff variation is 87.}

\colfig{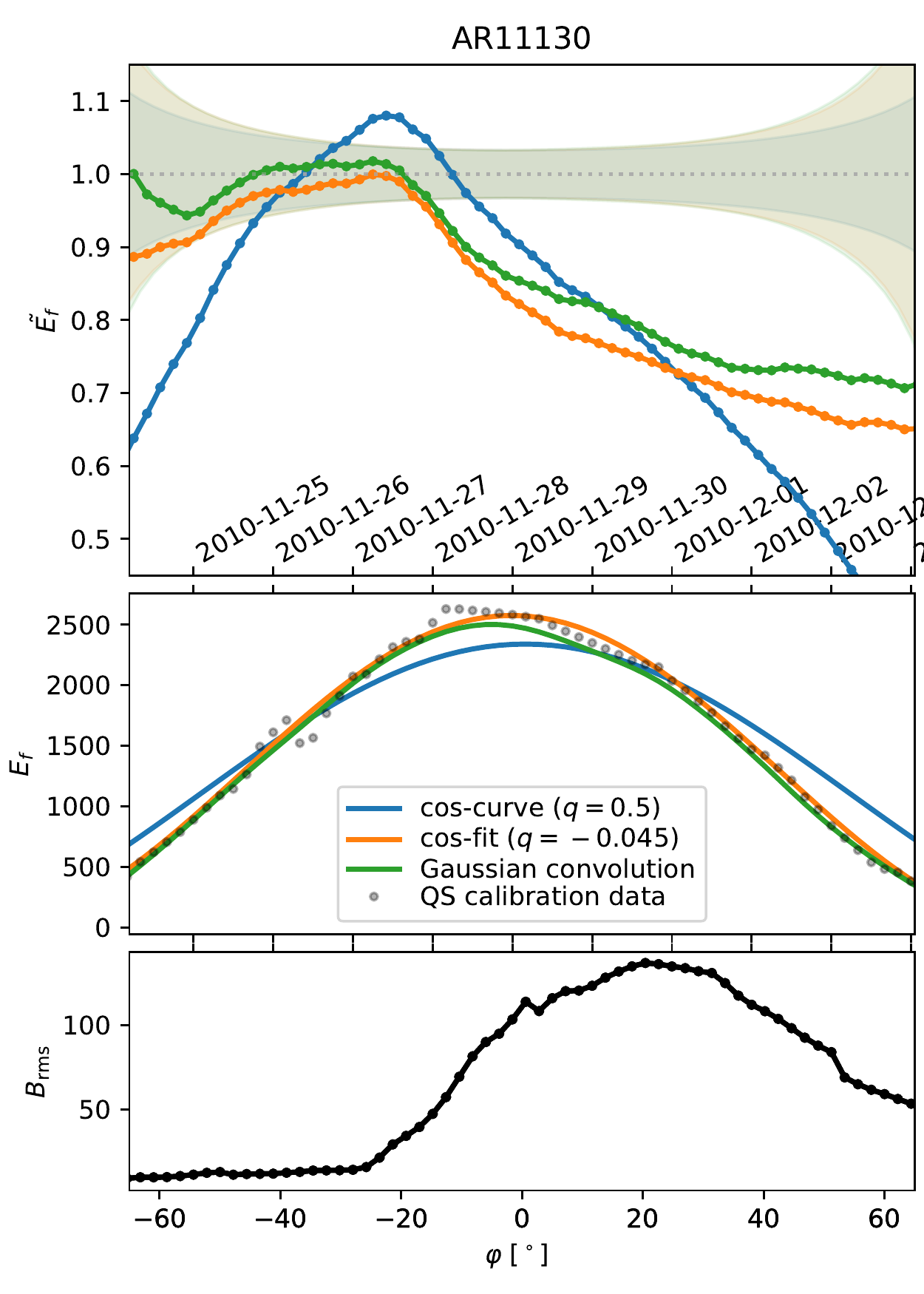}{Calibrated \fff strength (\eft, top), QS calibration data (\ef, gray dots) and fits (middle) and \brms\ (bottom) for AR11130. The line-plot colors in the top and the middle panel indicate \eft and \ef for various calibration methods as indicated in the legend. The shaded area in the top panel represents the standard deviation of the QS calibration derived from \fig{fmode_vs_brms} and scaled with the fitted calibration functions.}
\colfig{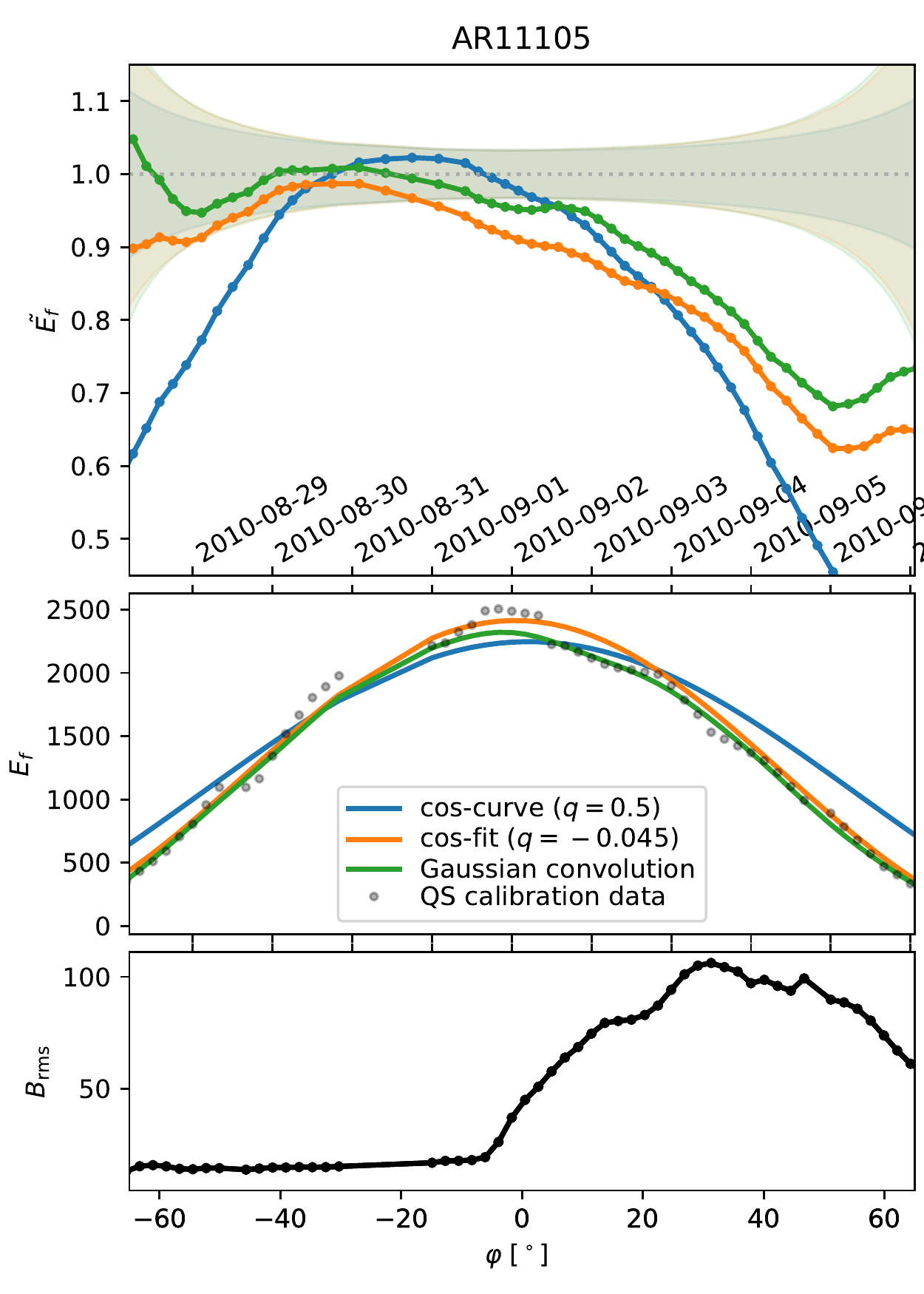}{Same as \fig{AR11130_track}, but for AR11105.}

\section{Discussion and conclusions}

The first part of the paper presented an analysis of QS magnetism, where the main findings were the solar cycle independence of the \IN{} magnetic fluctuations, studied by using $1\degree$ patches, while \NW{} and \IN{} fields, studied by using $15 \degree$ patches, were found to show a statistically significant solar cycle dependence. The maximum of the magnetic fluctuations at the disk center was observed to exhibit a phase shift of roughly half a year w.r.t. to the sunspot-number-defined cycle. We interpret these findings as follows. The independence of the \IN{} magnetic fields is either reflecting the fact that these fields, clearly on smaller scales than the supergranular cells, are generated by the fluctuation (or small-scale) dynamo alone, or then the sensitivity of the HMI instrument prevents one to see any variation. The solar cycle dependence of the \NW{} and \IN{} fields reflects the fact that two separate processes are responsible for generating these fields. Turbulence is tangling the solar-cycle dependent large-scale sub-surface field resulting in magnetic fluctuations that have the same dependence, and these are then expelled to the edges of the supergranular cells. The magnetic fluctuations generated by the fluctuation dynamo also become expelled in a similar fashion, but as the growth of these fluctuations is exponential due to the dynamo instability, the fluctuation dynamo can quickly replenish the fluctuations also in the \IN{} regions. The tangling process can be envisioned to have an amplification time scale that is slower, only linear over time.

As is extensively reviewed in \cite{BS05}, in turbulent dynamo theory, the magnetic fields grow under the constraint of magnetic helicity
conservation, which leads to a bihelical spectrum of magnetic helicity. Signs of such bihelical nature of the solar surface magnetic field have already been reported by \citet{B+17,S+18}. Accumulation of small-scale
magnetic helicity leads to the quenching of large-scale dynamo. In order to further grow its large-scale
magnetic field, the system must shed its small-scale helicity. Being an open system, the Sun
may have fluxes of magnetic helicity where ARs could play a vital role in removing the magnetic
helicity from small-scales, thus leading to a rejuvenation of
the
large-scale dynamo. This mechanism was proposed to lead to a phase shift 
between the cycles for sunspot and magnetic helicity,
that was seen to peak half a year later than the sunspot cycle in \cite{S+18}. We see similar evidence here for the solar-cycle dependent part of the magnetic fluctuations in the \NW{} and \IN{} - the phase shift by half a year is consistent with the large-scale field being rejuvenated later, when ARs have already peaked, and shed some small-scale magnetic helicity out from the system to allow for the growth of the global field.

The second part of our study built a novel method for calibrating the \fff energy using the very quiet patches harvested in the first part. Even though more than 22\,000 patches were analysed to obtain the statistics over the solar cycle, the \fff energy variations of the QS remained large, and they show no clear correlation with the 
rms magnetic field strength
measured at the surface. We interpret this as an indication of the \fff energy being affected by sub-surface magnetic fields, and the surface rms magnetic field being an inadequate indicator to find the QS level. 
The measured QS \fff energy, indeed, shows a solar-cycle variation, which is, in broad terms, anti-correlated with the solar cycle, the \fff energy being lower/higher when the
rms
magnetic field is strong/weak.
Such a behavior is expected in the light of many previous investigations, which have suggested \fff damping to occur as a result of absorption of the \fff mode by the magnetic fields in ARs \citep{Cally+94,CB97,SRB16}. This effect is the dominant one over the
possible enhancement prior to AR emergence,
as can be seen also in the hindcasted AR data in this 
study. Combined with the heavy averaging, any possible enhancement 
will be washed out from a statistic measuring the global behavior.

The phase shift of several years between the sunspot cycle and the \fff modulation is too large to be explicable by the scenario suggested above for the half a year phase shift of the QS magnetism maximum w.r.t. the sunspot cycle. The \fff energy modulation appears to follow more closely the poloidal component of the global field than the toroidal part that is commonly thought to give rise to the ARs. These two field components are known to have a systematic phase difference of roughly $\pi$ \cite[see, e.g.,][]{char10}. 

The general trend of \fff energy being reduced during SC25 in comparison to SC24 could be a result of the instrument aging, but the clear differences seen over different latitudes and hemispheres is speaking against of an overall instrumental degradation. If it were a physical effect, it would suggest that the sub-surface magnetic field is stronger during the ascending phase of SC25 than the one of SC24, and that could lead to SC25 being stronger than SC24. Further investigations and fine tuning of the method are required to verify the results and conclusions reached in this study, but the QS \fff energy seems to have potential applications in diagnosing the sub-surface magnetic fields in the Sun.

The third and last part of our study presented two hindcasted 
AR \fff signals, that have been proposed to show a clear enhancement of the \fff energy in the earlier studies \citep[][]{SRB16,Waidele22}, with our novel QS calibration method. In summary, our analysis shows that the transient
and weak enhancement signal is very sensitive to the calibration
method used. The new method for the QS calibration developed here tends to show a more flat signal with mild enhancement only in comparison to the earlier calibration method of \cite{SRB16}. Although in general a signature of an enhancement is seen, it remains unclear whether it is an effect caused by the uncertainties close to the limb or a real effect. In the light of the QS data, the magnitude of the enhancement is weaker than the QS variability level, and hence its detection and usage as an AR emergence predictor is not possible with the QS calibration method.

\begin{acknowledgements}
We are indebted to Dr. Harsha Raichur, the developer of the original \fff analysis pipeline, wherefrom this work stems. We acknowledge the fruitful discussions with Prof. Nishant Singh on the theoretical interpretation of the results and practical insights to the analysis.
All SDO data used are publicly available from the Joint Science Operations Center (JSOC) at Stanford University supported by NASA Contract NAS5- 02139 (HMI), see http://jsoc.stanford.edu/. 
The data analysis has been carried out on supercomputers in the facilities hosted by the CSC---IT
Center for Science in Espoo, Finland, which are financed by the
Finnish ministry of education.
The data were also
processed at the German Data Center for SDO (GDC-SDO), funded by the
German Aerospace Center (DLR), and hosted by the Max Planck Institute for Solar System Research (Göttingen, Germany).
This project has received funding from the European Research Council (ERC)
under the European Union's Horizon 2020 research and innovation
program (Project UniSDyn, grant agreement n:o 818665).

\end{acknowledgements}

\bibliography{main}{}
\bibliographystyle{aasjournal}

\end{document}